\documentclass[12pt]{article}
\usepackage{graphicx,epsfig}
\usepackage{epsf}
\usepackage[table]{xcolor}
\usepackage{colortbl}
\definecolor{lightgray}{gray}{0.9}
\usepackage{graphicx}
\usepackage{lscape}
\usepackage{citesort}
\usepackage{amssymb}
\usepackage{appendix}
\usepackage{multirow}

\begin{document}
\def\qq{\langle \bar q q \rangle}
\def\uu{\langle \bar u u \rangle}
\def\dd{\langle \bar d d \rangle}
\def\sp{\langle \bar s s \rangle}
\def\GG{\langle g_s^2 G^2 \rangle}
\def\Tr{\mbox{Tr}}
\def\figt#1#2#3{
        \begin{figure}
        $\left. \right.$
        \vspace*{-2cm}
        \begin{center}
        \includegraphics[width=10cm]{#1}
        \end{center}
        \vspace*{-0.2cm}
        \caption{#3}
        \label{#2}
        \end{figure}
    }

\def\figb#1#2#3{
        \begin{figure}
        $\left. \right.$
        \vspace*{-1cm}
        \begin{center}
        \includegraphics[width=10cm]{#1}
        \end{center}
        \vspace*{-0.2cm}
        \caption{#3}
        \label{#2}
        \end{figure}
                }

\def\ds{\displaystyle}
\def\beq{\begin{equation}}
\def\eeq{\end{equation}}
\def\bea{\begin{eqnarray}}
\def\eea{\end{eqnarray}}
\def\beeq{\begin{eqnarray}}
\def\eeeq{\end{eqnarray}}
\def\ve{\vert}
\def\vel{\left|}
\def\ver{\right|}
\def\nnb{\nonumber}
\def\ga{\left(}
\def\dr{\right)}
\def\aga{\left\{}
\def\adr{\right\}}
\def\lla{\left<}
\def\rra{\right>}
\def\rar{\rightarrow}
\def\lrar{\leftrightarrow}
\def\nnb{\nonumber}
\def\la{\langle}
\def\ra{\rangle}
\def\ba{\begin{array}}
\def\ea{\end{array}}
\def\tr{\mbox{Tr}}
\def\ssp{{\Sigma^{*+}}}
\def\sso{{\Sigma^{*0}}}
\def\ssm{{\Sigma^{*-}}}
\def\xis0{{\Xi^{*0}}}
\def\xism{{\Xi^{*-}}}
\def\qs{\la \bar s s \ra}
\def\qu{\la \bar u u \ra}
\def\qd{\la \bar d d \ra}
\def\qq{\la \bar q q \ra}
\def\gGgG{\la g^2 G^2 \ra}
\def\q{\gamma_5 \not\!q}
\def\x{\gamma_5 \not\!x}
\def\g5{\gamma_5}
\def\sb{S_Q^{cf}}
\def\sd{S_d^{be}}
\def\su{S_u^{ad}}
\def\sbp{{S}_Q^{'cf}}
\def\sdp{{S}_d^{'be}}
\def\sup{{S}_u^{'ad}}
\def\ssp{{S}_s^{'??}}

\def\sig{\sigma_{\mu \nu} \gamma_5 p^\mu q^\nu}
\def\fo{f_0(\frac{s_0}{M^2})}
\def\ffi{f_1(\frac{s_0}{M^2})}
\def\fii{f_2(\frac{s_0}{M^2})}
\def\O{{\cal O}}
\def\sl{{\Sigma^0 \Lambda}}
\def\es{\!\!\! &=& \!\!\!}
\def\ap{\!\!\! &\approx& \!\!\!}
\def\md{\!\!\!\! &\mid& \!\!\!\!}
\def\ar{&+& \!\!\!}
\def\ek{&-& \!\!\!}
\def\kek{\!\!\!&-& \!\!\!}
\def\cp{&\times& \!\!\!}
\def\se{\!\!\! &\simeq& \!\!\!}
\def\eqv{&\equiv& \!\!\!}
\def\kpm{&\pm& \!\!\!}
\def\kmp{&\mp& \!\!\!}
\def\mcdot{\!\cdot\!}
\def\erar{&\rightarrow&}
\def\olra{\stackrel{\leftrightarrow}}
\def\ola{\stackrel{\leftarrow}}
\def\ora{\stackrel{\rightarrow}}

\def\simlt{\stackrel{<}{{}_\sim}}
\def\simgt{\stackrel{>}{{}_\sim}}


\title{
         {\Large
                 {\bf
                       Semileptonic $B_{s} \rightarrow D_{s2}^{*}(2573)\ell\bar{\nu}_{\ell}$   transition in QCD
                 }
         }
      }

\author{\vspace{1cm}\\
{\small K. Azizi$^1$ \thanks
{e-mail: kazizi@dogus.edu.tr }\,\,,\,\,H. Sundu$^2$ \thanks {e-mail: hayriye.sundu@kocaeli.edu.tr}\,\,,\,\,
S. \c{S}ahin$^2$ \thanks {e-mail: 095131004@kocaeli.edu.tr}}  \\
{\small $^1$ Department of Physics, Do\u gu\c s University,
Ac{\i}badem-Kad{\i}k\"oy, 34722 Istanbul, Turkey }\\
{\small $^2$ Department of Physics , Kocaeli University, 41380
Izmit, Turkey} }
\date{}

\begin{titlepage}
\maketitle
\thispagestyle{empty}

\begin{abstract}
We analyze the semileptonic $ B_{s} \rightarrow
D_{s2}^{*}(2573)\ell\bar{\nu }_{\ell}$ transition, where $\ell=\tau,~\mu$ or $e$,
 within the standard model. We apply the QCD sum rule approach to calculate the transition form factors entering the low energy Hamiltonian defining this channel. The fit functions of the form factors
are used to estimate the total decay  widths and  branching fractions in all lepton channels. The orders of branching ratios  indicate that this transition is accessible at LHCb
in near future.
\end{abstract}

~~~PACS number(s): 13.20.-v, 13.20.He,  11.55.Hx
\end{titlepage}

\section{Introduction}

The semileptonic $B$ meson decay channels are known as  useful tools to
accurately calculate the Standard Model (SM) parameters like determination of the Cabibbo-Kobayashi-Maskawa (CKM) quark mixing matrix, check the validity of the SM,  describe the origin of the CP
violation and search for new physics effects.
By recent experimental progresses, it has become precise  measurements available, and it is possible to
perform precision calculations. Although the  $B$
meson decays are studied efficiently both theoretically and
experimentally (see for instance \cite{amo,hok,san,aaij,leesj,boz,faj,colangelo,Becirevic,elvir,yang}),
most of $B_{s}$ properties are not very clear yet (for some related theoretical and experimental studies on this meson see \cite{lees,aj,aij,abaz,eric,gan,Albertus,Steinhauser,Bouchard,Albertus1} and references therein).
Since the  detection and identification of this heavy meson is relatively
difficult in the experiment,  the theoretical and phenomenological
studies on the spectroscopy and decay properties of this mesons can play essential role in our
understanding of its  non-perturbative dynamics, calculating the related parameters of the SM and providing opportunities to search for
possible  new physics contributions.


 In the literature, there are a lot of theoretical studies devoted to the semileptonic transition of $ B_{s}$ into the pseudoscalar $D_{s}$ and vector  $D^*_{s}$ charmed-strange mesons. But, we have no study on the
semileptonic transitions of this meson into the tensor charmed-strange meson in final state, although it is expected to have considerable contribution to the total decay width of the $ B_{s}$ meson.
In this accordance, in the present study, we investigate  the semileptonic $ B_{s} \rightarrow D_{s2}^{*}(2573)\ell\bar{\nu }_{\ell}$ transition in the framework of three-point QCD sum rule  \cite{shif}
as one of the most attractive and powerful techniques in hadron phenomenology, where the $D_{s2}^{*}(2573)$ is the low lying charmed-strange tensor meson with $J^P=2^+$. In particular,
we calculate the transition form factors entering the low energy matrix elements defining the transition under consideration. We find the working regions of the auxiliary parameters entering
the calculations from different transformations, considering the criteria of the method used. This is followed by finding the behavior of the form factors in terms of the transferred momentum squared,
 which are then used to estimate the total width and branching fraction in all lepton channels. Note that the semileptonic
 $ B\rightarrow D_{2}^{*}(2460)\ell\bar{\nu }_{\ell}$ decay channel is analyzed in \cite{sahin} using the same method. The spectroscopic properties of the charmed-strange tensor meson $D_{s2}^{*}(2573)$ is also investigated in
\cite{azizi} using a two-point correlation function.




The layout of the paper is as follows.  In next section, the QCD
sum rules for the four form factors relevant to the semileptonic $
B_{s} \rightarrow D_{s2}^{*}(2573)\ell\bar{\nu }_{\ell}$ transition are
obtained.  Section 3 contains numerical analysis of the form
factors, calculation of their behavior in terms of $q^2$ as well as the estimation of the total decay width and branching ratio for the transition under consideration.

\section{Theoretical framework}

In order to calculate the form factors, associated with the semileptonic $ B_{s}
\rightarrow D_{s2}^{*}(2573)\ell\bar{\nu }_{\ell}$ transition via QCD sum rule formalism, we consider the following three-point correlation
function:

\begin{eqnarray}\label{correl.func.101}
\Pi _{\mu\alpha\beta}=i^2\int d^{4}x\int d^{4}y
e^{-ip.x}e^{ip'.y}{\langle}0\mid {\cal
T}\Big[J_{\alpha\beta}^{D_{s2}^{*}(2573)}(y) J^{tr}_{\mu}(0)
J^{B_{s}^{\dag}}(x)\Big]\mid0{\rangle},
\end{eqnarray}
where ${\cal T}$ is the
time ordering operator and
$J^{tr}_{\mu}(0)=\bar{c}(0)\gamma_{\mu}(1-\gamma_5)b(0)$ is the
transition current. The interpolating currents of the $B_{s}$ and
$D_{s2}^{*}(2573)$  mesons can be written
in terms of the quark fields as
\begin{eqnarray}\label{pseudscalarcurrent}
 J^{B_{s}}=\bar{s}(x)\gamma_5b(x),
\end{eqnarray}
and
\begin{eqnarray}\label{tensorcurrent}
J_{\alpha\beta}^{D_{s2}^{*}(2573)}(y)=\frac{i}{2}\left[\bar s(y)
\gamma_{\alpha} \olra{\cal D}_{\beta}(y) c(y)+\bar s(y)
\gamma_{\beta} \olra{\cal D}_{\alpha}(y) c(y)\right].
\end{eqnarray}
Here $ \olra{\cal D}_{\beta}(y)$ is the covariant derivative
that acts on the left and right, simultaneously.
It is given as
\begin{eqnarray}\label{derivative}
\olra{\cal D}_{\beta}(y)=\frac{1}{2}\left[\ora{\cal D}_{\beta}(y)-
\ola{\cal D}_{\beta}(y)\right],
\end{eqnarray}
with
\begin{eqnarray}\label{derivative2}
\overrightarrow{{\cal
D}}_{\beta}(y)=\overrightarrow{\partial}_{\beta}(y)-i
\frac{g}{2}\lambda^aA^a_\beta(y),\nonumber\\
\overleftarrow{{\cal
D}}_{\beta}(y)=\overleftarrow{\partial}_{\beta}(y)+
i\frac{g}{2}\lambda^aA^a_\beta(y),
\end{eqnarray}
where  $\lambda^a$ and $A^a_\beta(x)$ denote the Gell-Mann matrices and
the external  gluon fields, respectively.

According to the method used, in order to find the QCD sum rules for transition form
factors, we shall calculate the aforesaid correlation function, once in terms of hadronic parameters and the second in terms of QCD parameters making use of operator product expansion (OPE).
By equating these two representations to each other through a dispersion relation, we obtain the sum rules for form factors.  To
stamp down the contributions of the higher states and continuum,
a double Borel transformation with respect to the $p^{2}$ and
$p^{\prime^{2}}$ is performed on both sides of the sum rules obtained and the quark-hadron duality assumption is used.

\subsection{The hadronic representation}

In order to calculate the hadronic side of the correlator in Eq.(\ref{correl.func.101}),
 we insert two complete sets of the initial $B_{s}$ and the final $D_{s2}^{*}(2573)$ states
 with the same quantum numbers as the interpolating currents into the correlator. After performing
 four-integrals over $x$ and $y$, we obtain
\begin{eqnarray}\label{phen1}
\Pi _{\mu\alpha\beta}^{had}&=&\frac{{\langle}0\mid  J
_{\alpha\beta}^{D_{s2}^{*}(2573)}\mid
D_{s2}^{*}(2573)(p',\epsilon)\rangle{\langle}D_{s2}^{*}(2573)(p',\epsilon)\mid
J _{\mu}^{tr}(0) \mid B_{s}(p)\rangle \langle B_{s}(p)\mid
J_{B_{s}}^{\dag}\mid
 0\rangle}{(p^2-m_{B_{s}}^2)(p'^2-m_{D_{s2}^{*}(2573)}^2)}
 \nonumber \\
  &+& \cdots,\nonumber\\
\end{eqnarray}
 where $\cdots$ represents  contributions of  the higher states and continuum, and
 $\epsilon$ is the polarization tensor of the $D_{s2}^{*}(2573)$ tensor meson. We can parameterize the matrix
elements appearing in the above equation in terms of decay
constants, masses and form factors as
\begin{eqnarray}\label{lep}
{\langle}0\mid  J_{\alpha\beta}^{D_{s2}^{*}(2573)}\mid
D_{s2}^{*}(2573)(p',\epsilon)\rangle&=&m_{D_{s2}^{*}(2573)}^3
f_{D_{s2}^{*}(2573)}\epsilon_{\alpha\beta},\nonumber \\
\langle B_{s}(p)\mid
J_{B_{s}}^{\dag}\mid
 0\rangle&=&-i\frac{f_{B_{s}} m_{B_{s}}^2} {m_s+m_b},
\nonumber \\
{\langle}D_{s2}^{*}(2573)(p',\epsilon)\mid J
_{\mu}^{tr}(0) \mid B_{s}(p)\rangle&=&h(q^2)\varepsilon_{\mu\nu\alpha\beta}\epsilon^{*^{\nu\lambda}}P^{\lambda}
P_{\alpha}q^{\beta}-iK(q^2)\epsilon^{*_{\mu\nu}}P^{\nu}\nonumber \\
&-&i\epsilon^{*_{\alpha\beta}}P^{\alpha}P^{\beta}\left[P_{\mu}b_{+}(q^2)+q_{\mu}b_{-}(q^2)
\right],
\end{eqnarray}
where $q=p-p'$, $P=p+p'$; and $h(q^2)$, $K(q^2)$, $b_{+}(q^2)$ and $b_{-}(q^2)$ are transition form factors.
 Now, we  combine Eqs. (\ref{phen1})
and  (\ref{lep}) and performing summation over the  polarization
tensors via
\begin{eqnarray}\label{polarizationt1}
\epsilon_{\alpha\beta}\epsilon_{\nu\theta}^*=\frac{1}{2}T_{\alpha\nu}T_{\beta\theta}+
\frac{1}{2}T_{\alpha\theta}T_{\beta\nu}
-\frac{1}{3}T_{\alpha\beta}T_{\nu\theta},
\end{eqnarray}
where
\begin{eqnarray}\label{polarizationt2}
T_{\alpha\nu}=-g_{\alpha\nu}+\frac{p'_\alpha
p'_\nu}{m_{D_{s2}^{*}(2573)}^2}.
\end{eqnarray}
This procedure brings us to the final representation of the hadronic side, viz.
\begin{eqnarray}\label{phen2}
\Pi _{\mu\alpha\beta}^{had}&=&\frac{f_{D_{s2}^{*}}f_{B_{s}}
m_{D_{s2}^{*}} m_{B_{s}}^2}
{8(m_b+m_s)(p^2-m^2_{B_{s}})(p'^2-m_{D_{s2}^{*}}^2)}
\Bigg\{\frac{2}{3}\Big[-\Delta K(q^2)+\Delta'b_{-}(q^2)
\Big]q_{\mu}g_{\beta\alpha}\nonumber \\
 &+&\frac{2}{3}\Big[(\Delta-4m_{D_{s2}^{*}}^2) K(q^2)+\Delta'b_{+}(q^2)
\Big]P_{\mu}g_{\beta\alpha}+i(\Delta-4m_{D_{s2}^{*}}^2)
h(q^2)\varepsilon_{\lambda\nu\beta\mu}P_{\lambda}P_{\alpha}q_{\nu}
\nonumber \\
&+&\Delta K(q^2)q_{\alpha}g_{\beta\mu}+ \mbox{other
structures}\Bigg\}+...,
\end{eqnarray}
where
\begin{eqnarray}\label{phen2}
\Delta &=&m_{B_{s}}^2+3m_{D_{s2}^{*}}^2-q^2,
\end{eqnarray}
and
\begin{eqnarray}
\Delta'&=&m_{B_{s}}^4-2m_{B_{s}}^2(m_{D_{s2}^{*}}^2+q^2)+(m_{D_{s2}^{*}}^2-q^2)^2.
\end{eqnarray}

\subsection{The OPE representation}

The OPE  side of the correlation function is calculated in deep Euclidean region. For this aim, we insert the explicit forms of the interpolating currents into
  the correlation function in Eq. (\ref{correl.func.101}). After performing contractions via the Wick's theorem, we obtain the following result in terms of the heavy and light quarks propagators:
\begin{eqnarray}\label{correl.func.2}
\Pi^{OPE} _{\mu\alpha\beta}&=&\frac{-i^3}{2}\int d^{4}x\int
d^{4}ye^{-ip\cdot x}e^{ip'\cdot y}
\nonumber \\
&\times& \Bigg\{Tr\left[S_s^{ki}(x-y)\gamma_\alpha\olra{\cal
D}_{\beta}(y)
S_c^{ij}(y)\gamma_\mu(1-\gamma_5)S_b(-x)^{jk}\gamma_5\right]+
\left[\beta\leftrightarrow\alpha\right]
\Bigg\}.\nonumber\\
\end{eqnarray}
The heavy and light quarks propagators appearing in above equation and up to terms taken into account in the calculations are given by
\begin{eqnarray}\label{heavypropagator}
S_{Q}^{i\ell}(x)&=&\frac{i}{(2\pi)^4}\int d^4k e^{-ik \cdot x}
\left\{ \frac{\delta_{i\ell}}{\!\not\!{k}-m_Q}
-\frac{g_sG^{\alpha\beta}_{i\ell}}{4}\frac{\sigma_{\alpha\beta}(\!\not\!{k}+m_Q)+
(\!\not\!{k}+m_Q)\sigma_{\alpha\beta}}{(k^2-m_Q^2)^2}\right.\nonumber\\
&&\left.+\delta_{i\ell}\frac{\pi^2}{3} \langle
\frac{\alpha_sGG}{\pi}\rangle \frac{
m_Qk^2+m_Q^2\!\not\!{k}}{(k^2-m_Q^2)^4} +\cdots\right\} \, ,
\end{eqnarray}
where $Q=b$ or $c$,
and
\begin{eqnarray}\label{lightpropagator}
S_{s}^{ij}(x)&=& i\frac{\!\not\!{x}}{ 2\pi^2 x^4}\delta_{ij}
-\frac{m_s}{4\pi^2x^2}\delta_{ij}-\frac{\langle
\bar{s}s\rangle}{12}\Big(1 -i\frac{m_s}{4}
\!\not\!{x}\Big)\delta_{ij} -\frac{x^2}{192}m_0^2\langle
\bar{s}s\rangle\Big(1-i\frac{m_s}{6} \!\not\!{x}\Big)\delta_{ij}
\nonumber \\
&-&\frac{ig_s
G_{\theta\eta}^{ij}}{32\pi^2x^2}\big[\!\not\!{x}\sigma^{\theta\eta}+\sigma^{\theta\eta}\!\not\!{x}\big]
+\cdots \, .
\end{eqnarray}

To proceed, we insert the expressions of the heavy and light propagators into Eq. (\ref{correl.func.2}) and perform the  derivatives
with respect to $x$ and $y$. Then, we transform the calculations  to the momentum space
 and make the $x_{\mu}\rightarrow
i\frac{\partial}{\partial p_{\mu}}$ and
 $y_{\mu}\rightarrow -i\frac{\partial}{\partial p'_{\mu}}$ replacements. We perform the two four-integrals  coming from the heavy quark propagators with the help of
 two Dirac delta functions appearing in the calculations. Finally, we perform the last four-integral  using the Feynman parametrization, viz.

\begin{eqnarray}\label{Int}
\int d^4t\frac{(t^2)^{\beta}}{(t^2+L)^{\alpha}}=\frac{i \pi^2
(-1)^{\beta-\alpha}\Gamma(\beta+2)\Gamma(\alpha-\beta-2)}{\Gamma(2)
\Gamma(\alpha)[-L]^{\alpha-\beta-2}}.
\end{eqnarray}
Eventually, we get the OPE side of the three-point correlation function in terms of the selected structures and the perturbative and non-perturbative parts as
\begin{eqnarray}\label{QCDside}
\Pi^{OPE}_{\mu\alpha\beta}&=&\Big(\Pi^{pert}_1(q^2)+\Pi^{non-pert}_1(q^2)\Big)q_{\alpha}g_{\beta\mu}+
\Big(\Pi^{pert}_2(q^2)+\Pi^{non-pert}_2(q^2)\Big)q_{\mu}g_{\beta\alpha}\nonumber \\
&+&
\Big(\Pi^{pert}_3(q^2)+\Pi^{non-pert}_3(q^2)\Big)P_{\mu}g_{\beta\alpha}+
\Big(\Pi^{pert}_4(q^2)+\Pi^{non-pert}_4(q^2)\Big)\varepsilon_{\lambda\nu\beta\mu}P_{\lambda}
P_{\alpha}q_{\nu}
\nonumber \\
&+&\mbox{other structures},
\end{eqnarray}
where the perturbative parts
$\Pi^{pert}_i(q^2)$  can be written in terms of the double dispersion
integrals as
\begin{eqnarray}\label{QCDside}
\Pi^{pert}_i(q^2)=\int^{}_{}ds\int^{}_{}ds'
\frac{\rho_i(s,s',q^2)}{(s-p^2)(s'-p'^2)}.
\end{eqnarray}
The $O(1)$ spectral densities $\rho_i(s,s',q^2)$ are
given by the imaginary parts of the $\Pi^{pert}_{i}(q^2)$
functions, i.e.,
$\rho_i(s,s',q^2)=\frac{1}{\pi}Im[\Pi^{pert}_{i}(q^2)]$. After lengthy calculations, the  spectral densities corresponding to the selected structures are obtained as
\begin{figure}[h!]
\begin{center}
\includegraphics[width=15cm]{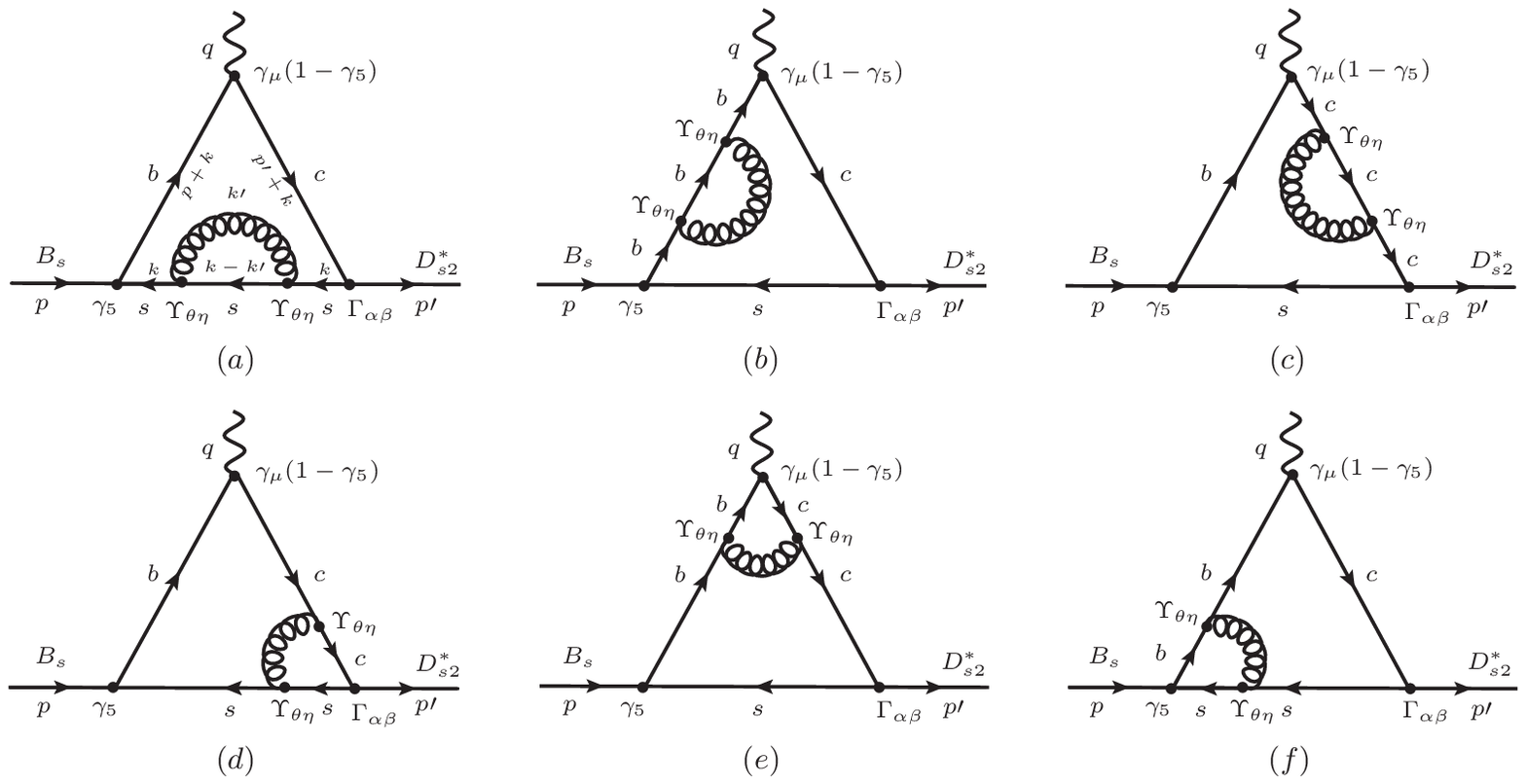}
\end{center}
\caption{Perturbative ${\cal O}(\alpha_s)$  diagrams contributing to the correlation function.} \label{Graphs}
\end{figure}

\begin{eqnarray}\label{rho}
\rho_1(s,s',q^2)&=&\int_{0}^{1}dx\int^{1-x}_{0} dy
\Bigg\{{-}\frac{3 (m_c(4+6y-6x)+m_s(-2+y-x)+2 m_b
(y-x))}{64\pi^2}\Bigg\}\nonumber\\&\times&\Theta[L(s,s',q^2)],
\nonumber \\
\rho_2(s,s',q^2)&=&\int_{0}^{1}dx\int^{1-x}_{0} dy
\Bigg\{\frac{3((y-x)(m_b x + m_c
(-1+2y+x))}{32\pi^2(-1+y+x)}\Bigg\}\Theta[L(s,s',q^2)],
\nonumber \\
\rho_3(s,s',q^2)&=&\int_{0}^{1}dx\int^{1-x}_{0} dy
\Bigg\{{-}\frac{3 (m_b x
(-2+3y+x)+m_c(2y^2+y(-1+x)+(-1+x)x))}{32\pi^2(-1+y+x)}\Bigg\}\nonumber\\&\times&\Theta[L(s,s',q^2)],
\nonumber \\
\rho_4(s,s',q^2)&=&0,
\end{eqnarray}
where $\Theta[...]$ is the unit-step function and
\begin{eqnarray}\label{L}
L(s,s',q^2)=-m_c^2y-s'y(x+y-1)-x\Big(m_b^2-q^2y+s(x+y-1)\Big).
\end{eqnarray}

We also take into account the perturbative ${\cal O}(\alpha_s)$  corrections contributing to the correlation function. These corrections for massless quarks 
   are  calculated using the standard Cutkosky rules in \cite{Braguta}  for calculation of pion form factor with both 
the pseudoscalar and axial currents. These corrections are also calculated in  the case of  transition between  two infinitely heavy quarks with the spectator quark being massless in \cite{colangelobey} using the
 universal Isgur-Wise function. We calculate the ${\cal O}(\alpha_s)$  corrections  keeping also the spectator strange quark mass in the calculations. 
 For this aim we consider the diagrams presented in figure 1. As an example
we present the amplitude of diagram $(a)$ in figure 1 which is obtained as
\begin{eqnarray}\label{rhoalfs}
\Pi_{\alpha_{s_{(a)}}}&=&-16\pi\alpha_s\int\frac{d^4k}{(2\pi)^4}\int\frac{d^4k^{\prime}}{(2\pi)^4}
\nonumber \\
&\times&\frac{Tr\Big[\Gamma_{\alpha\beta}(\!\not\!{p^{\prime}}+\!\not\!{k}+m_c)\gamma_{\mu}(1-\gamma_5)
(\!\not\!{p}+\!\not\!{k}+m_b)\gamma_5(\!\not\!{k}+m_s)\gamma^{\eta}(\!\not\!{k}-\!\not\!{k^{\prime}}+m_s)
\gamma^{\eta}(\!\not\!{k}+m_s)\Big]}{[(p^{\prime^2}+k)^2-m_c^2][(p+k)^2-m_b^2]
[(k-k^{\prime})^2-m_s^2](k^2-m_s^2)^2k^{\prime^2}},
\nonumber \\
\end{eqnarray}
where
\begin{eqnarray}\label{vertex}
\Gamma_{\alpha\beta}=\gamma_{\alpha}\left(2k_{\beta}+p^{\prime}_{\beta}\right)
+\gamma_{\beta}\left(2k_{\alpha}+p^{\prime}_{\alpha}\right)-\frac{2}{3}\left(g_{\alpha\beta}
-\frac{p^{\prime}_{\alpha}p^{\prime}_{\beta}}{p^{\prime^2}}\right)(2 \!\not\!{k}+\!\not\!{p'}).
\end{eqnarray}
After calculation of the four-integrals appearing in the
amplitudes of all diagrams shown in figure 1 and taking the
imaginary parts of the obtained results we select the
above-mentioned structures to find the ${\cal O}(\alpha_s)$
spectral densities $\rho_{\alpha_{s_i}}(s,s',q^2)$. The details of
calculations  for $\rho_{\alpha_{s_1}}(s,s',q^2)$ are given in appendix A.


The $\Pi^{non-pert}_i(q^2)$  functions are obtained up to five
dimension operators. As they have also very lengthy expressions, we do
not show their explicit form again.

%


%
%
%
%
Having calculated both the hadronic and OPE sides of the correlation function, we match the coefficients of the selected
structures from both sides and apply a double-Borel transformation. As a result, we get the following sum rules for the form factors:
\begin{eqnarray}\label{K}
K(q^2)&=&\frac{8(m_b+m_s)}{\Delta}\frac{1}{f_{B_s} f_{D_{s2}^{*}}
m_{D_{s2}^{*}} m_{B_s}^2}e^{\frac{m_{B_s}^2}{M^2}}
e^{\frac{m_{D_{s2}^{*}}^2}{M'^2}}\nonumber
\\
&\times&\Bigg\{\int^{s_0}_{(m_b+m_s)^2}ds
\int^{s_{0}^{'}}_{(m_c+m_s)^2}ds'\left(\rho_{1}(s,s',q^2)+\rho_{\alpha_{s_1}}(s,s',q^2)\right)e^{-\frac{s}{M^2}}e^{-\frac{s^{\prime}}{M^{\prime^2}}}
+\hat{B}\Pi_{1}^{non-pert}\Bigg\},
\nonumber \\
b_{-}(q^2)&=&\frac{12(m_b+m_s)}{f_{B_s} f_{D_{s2}^{*}}
m_{D_{s2}^{*}} m_{B_s}^2\Delta'} e^{\frac{m_{B_s}^2}{M^2}}
e^{\frac{m_{D_{s2}^{*}}^2}{M'^2}}
\nonumber \\
&\times& \Bigg\{\int^{s_0}_{(m_b+m_s)^2}ds
\int^{s_{0}^{'}}_{(m_c+m_s)^2}ds'\left(\rho_{2}(s,s',q^2)+\rho_{\alpha_{s_2}}(s,s',q^2)\right)e^{-\frac{s}{M^2}}e^{-\frac{s^{\prime}}{M^{\prime^2}}}
+\hat{B}\Pi_{2}^{non-pert}\Bigg\} \nonumber
\\
&+&\frac{\Delta}{\Delta'}K(q^2),
\nonumber \\
b_{+}(q^2)&=&\frac{12(m_b+m_s)}{f_{B_s} f_{D_{s2}^{*}}
m_{D_{s2}^{*}} m_{B_s}^2 \Delta'}e^{\frac{m_{B_s}^2}{M^2}}
e^{\frac{m_{D_{s2}^{*}}^2}{M'^2}}
\nonumber \\
&\times& \Bigg\{\int^{s_0}_{(m_b+m_s)^2}ds
\int^{s_{0}^{'}}_{(m_c+m_s)^2}ds'\left(\rho_{3}(s,s',q^2)+\rho_{\alpha_{s_3}}(s,s',q^2)\right)e^{-\frac{s}{M^2}}e^{-\frac{s^{\prime}}{M^{\prime^2}}}
+\hat{B}\Pi_{3}^{non-pert}\Bigg\}
\nonumber \\
&-&\frac{\Delta-4m^2_{D_{s2}^{*}}}{\Delta'}K(q^2),
\nonumber \\
h(q^2)&=&-i\frac{8(m_b+m_s)}{\Delta-4m^2_{D_{s2}^{*}}}\frac{1}{f_{B_s}
f_{D_{s2}^{*}} m_{D_{s2}^{*}} m_{B_s}^2}e^{\frac{m_{B_s}^2}{M^2}}
e^{\frac{m_{D_{s2}^{*}}^2}{M'^2}}\nonumber
\\
&\times& \Bigg\{\int^{s_0}_{(m_b+m_s)^2}ds
\int^{s_{0}^{'}}_{(m_c+m_s)^2}ds'\left(\rho_{4}(s,s',q^2)+\rho_{\alpha_{s_4}}(s,s',q^2)\right)e^{-\frac{s}{M^2}}e^{-\frac{s^{\prime}}{M^{\prime^2}}}
+ B \Pi_{4}^{non-pert}\Bigg\}, \nonumber
\\
\end{eqnarray}
where $M^2$ and $M^{'2}$ are the Borel mass parameters; and  $s_0$ and $s'_0$ are continuum thresholds in the initial and final mesonic channels, respectively.

\section{Numerical results}

In this section we present our numerical results for the transition form factors derived
from QCD sum rules and search for the behavior of the these quantities in terms of $q^2$.
To obtain numerical values, we use some input parameters presented in table 1.
\begin{table}[ht]\label{table1}
\centering \rowcolors{1}{lightgray}{white}
\begin{tabular}{cc}
\hline \hline
   Parameters  &  Values
           \\
\hline \hline
$ m_{B_{s}} $      &   $ (5366.77\pm0.24)  $  $ MeV $ \cite{olive}  \\
$ m_{D_{s2}^*(2573)}$    &   $ (2571.9\pm0.8) $ $MeV$ \cite{olive}  \\
$ f_{B_{s}} $      &   $ (222\pm12) $ $MeV$  \cite{baker} \\
$ f_{D_{s2}^*(2573)} $      &   $ (0.023\pm0.011) $ \cite{azizi} \\
$ G_{F} $            &   $ 1.17\times 10^{-5} $ $GeV^{-2}$ \\
$  V_{cb} $ &    $(41.2\pm1.1)\times 10^{-3}$   \\
$ \langle0|\overline{s}s|0\rangle$        &   $ -(0.8\pm0.24)^3$ $GeV^{3}$ \cite{reinders}  \\
$ m_0^2(1GeV) $       &   $ (0.8\pm0.2) $ $GeV^{2}$  \cite{reinders}  \\
$ \tau_{B_s} $       &   $(1.465\pm0.031)\times10^{-12}s$ \cite{olive}   \\
 \hline \hline
\end{tabular}
\caption{Input parameters used in  calculations.}
\end{table}

In our calculations, we also use the $\overline{MS}$ quark masses
$m_{c}(m_c)=(1.275\pm0.025)~GeV$, $m_{b}(m_b)=(4.18\pm0.03)~GeV$
and $m_s(\mu=2~GeV)=(95\pm5)~MeV$ \cite{olive}, and take into
account the energy-scale dependence of the $\overline{MS}$ masses
from the renormalization group equation to bring the masses to the same scale (see also \cite{Wang}),

\begin{eqnarray}\label{MassEqn}
m_b(\mu)&=&m_b(m_b)\left[\frac{\alpha_s(\mu)}{\alpha_s(m_b)}\right]^{\frac{12}{23}},
\nonumber \\
m_c(\mu)&=&m_c(m_c)\left[\frac{\alpha_s(\mu)}{\alpha_s(m_c)}\right]^{\frac{12}{25}},
\nonumber \\
m_s(\mu)&=&m_s(2~GeV)\left[\frac{\alpha_s(\mu)}{\alpha_s(2~GeV)}\right]^{\frac{4}{9}},
\end{eqnarray}
where
\begin{eqnarray}
 \alpha_s(\mu)&=&\frac{1}{b_0~t}\left[1-\frac{b_1}{b_0^2~t}\log[t]+\frac{b_1^2
\left(\log^2[t]-\log[t]-1\right)+b_0~b_2}{b_0^4~t^2}\right],
\end{eqnarray}
with
\begin{eqnarray}
\nonumber \\
b_2&=&\frac{1}{128~\pi^3}\left[2857-\frac{5033}{9}n_f+\frac{325}{27}n_f^2\right],
\nonumber \\
b_1&=&\frac{1}{24\pi^2}\left(153-19n_f\right),
\nonumber \\
b_0&=&\frac{1}{12\pi}\left(33-2n_f\right),
\nonumber \\
t&=&\log\left[\frac{\mu^2}{\Lambda^2}\right].
\end{eqnarray}
The parameter $\Lambda$  takes the values   $\Lambda=213~MeV$, $296~MeV$ and $339~MeV$ for the flavors
$n_f=5$, $4$ and $3$, respectively \cite{olive,Wang}.  We take 
$n_f=4$ in the present study. In \cite{Wang}  the authors take  $\mu=1~GeV$ for the  charmed and $\mu=3~GeV$ for the bottom tensor mesons. 
As we have the bottom and charmed mesons
respectively  in the initial and final states
in the transition under consideration, we take   the interval   $\mu=(2-4)~GeV$ for this parameter and discuss the rate of changes in the form factors and other observables when going from
 $\mu=2~GeV$ to $\mu=3~GeV$ and those from $\mu=3~GeV$ to  $\mu=4~GeV$.

To proceed further, we shall find  working regions of the  four
auxiliary parameters, namely the Borel mass parameters $M^2$ and
$M'^2$ and continuum thresholds $s_0$ and $s'_0$, such that the
transition form factors weakly depend on these parameters in those
regions. The continuum thresholds $s_0$ and $s'_0$ are the energy
squares which characterize the beginning of the continuum and
depend on the energy of the first excited states in the initial
and final channels, respectively.  Our numerical calculations
 point out the following regions for the continuum thresholds $s_0$ and $s'_0$:
 $29~GeV^2\leq s_0\leq35~GeV^2$ and $7~GeV^2\leq s'_0\leq11~GeV^2$.

The working regions for the Borel mass parameters are calculated demanding that both the higher
states and continuum are sufficiently suppressed and the contributions of the operators with
higher dimensions are small. As a result, we
find the working regions $10~GeV^2\leq M^2\leq 20GeV^2$ and $5GeV^2\leq M'^2\leq 10 GeV^2$ for Borel mass parameters.
To see  whether  the contributions related to the mesons of interest in the initial and  final states have been extracted by considering the above regions for the auxiliary parameters, we  calculate
the values of functions
$-d/d(1/M^2) \ln [\Pi^{OPE}(s_0,s'_0,M^2,M^{'2},q^2)]$ and $-d/d(1/M^{'2}) \ln[ \Pi^{OPE}(s_0,s'_0,M^2,M^{'2},q^2)]$ in the Borel scheme. Taking into account all the input parameters we find the
values $29.44~GeV^2\sim m_{B_s}^2$ and $5.58~GeV^2\sim m_{D_{s2}^{*}(2573)}^2$
for these functions, respectively, showing that  the contributions of the related mesons in the initial and  final states have been roughly extracted.
We show, as an example,  the dependence of    the form factor $K(q^2)$ at $q^2=1$
 on the Borel mass parameters $M^2$ and $M^{\prime^2}$ in figure 2. With a quick look at this figure, we see that not only this form factor depicts weak dependence on the
Borel parameters on their working regions, but the perturbative contribution constitutes the main part of the total value.

\begin{figure}[h!]
\begin{center}
\includegraphics[totalheight=7cm,width=7cm]{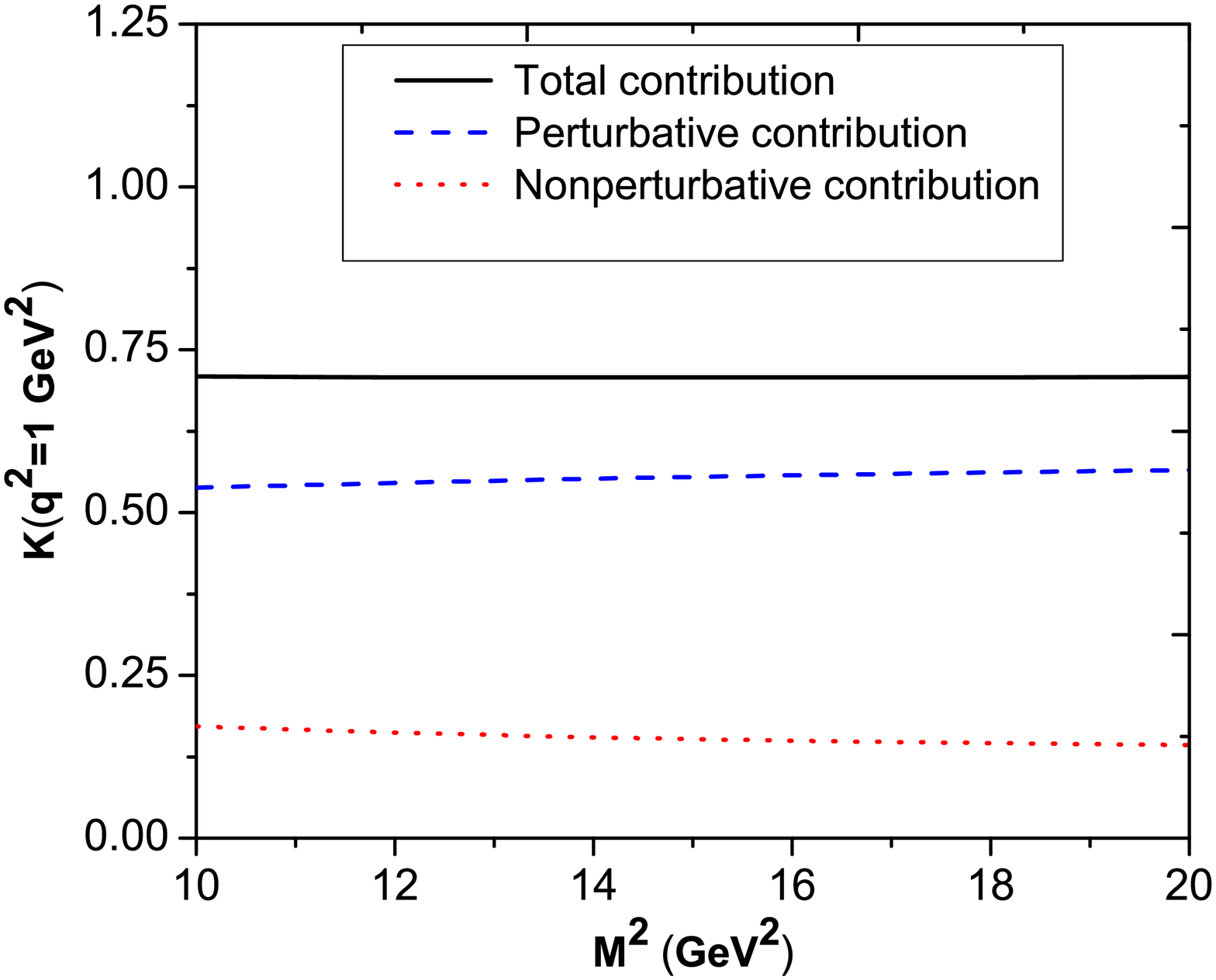}
\includegraphics[totalheight=7cm,width=7cm]{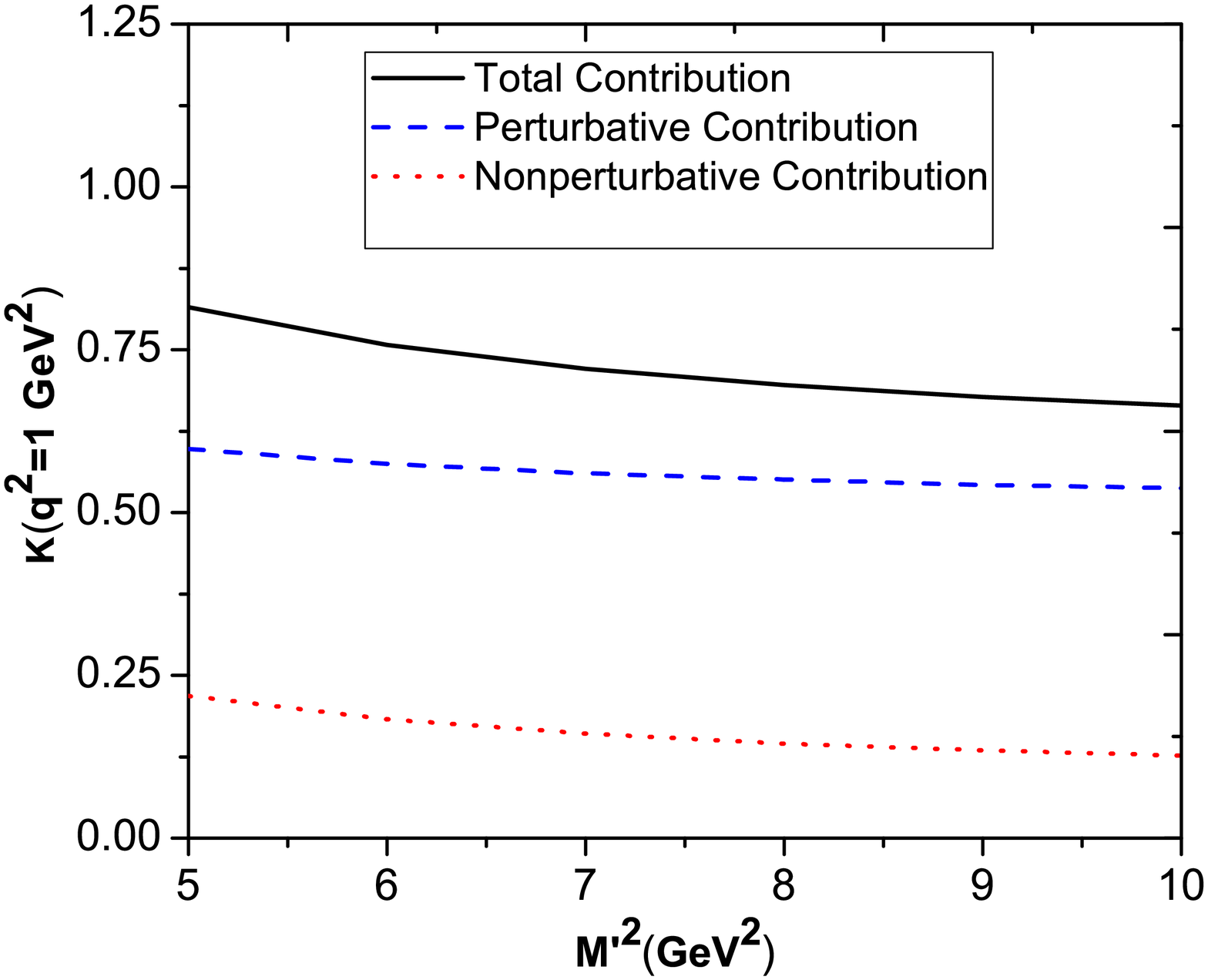}
\end{center}
\caption{\textbf{Left:} K($q^2=1$) as a function of the Borel mass
$M^2$ at average values of the $s_0$, $s'_0$ and $M^{\prime^2}$. \textbf{Right:}
 K($q^2=1$) as a function of the
Borel mass $M^{\prime^2}$ at average values of the $s_0$, $s'_0$ and $M^2$. } \label{KMsqMpsq}
\end{figure}

\begin{table}[h]
\renewcommand{\arraystretch}{1.5}
\addtolength{\arraycolsep}{3pt}
$$
\begin{array}{|c|c|c|c|c|}
\hline \hline
         &f_0 & \sigma_1 & \sigma_2   \\
\hline
  \mbox{$K (q^2)$} &0.70\pm0.30&-0.93\pm0.26&-1.93\pm0.58 \\
  \hline
  \mbox{$b_{-} (q^2)$} &(0.072\pm0.031)~GeV^{-2}&3.22\pm0.97&-1.72\pm0.82 \\
  \hline
  \mbox{$b_{+} (q^2)$} &(-0.031\pm0.013)~GeV^{-2}&4.07\pm1.22&1.39\pm0.41 \\
  \hline
  \mbox{$h (q^2)$} &(-0.0092\pm0.0038)~GeV^{-2}&0.33\pm0.10&-0.43\pm0.12 \\
                    \hline \hline
\end{array}
$$
\caption{Parameters appearing in the fit function of the form
factors at $\mu=2~GeV$.} \label{fitfunction1}
\renewcommand{\arraystretch}{1}
\addtolength{\arraycolsep}{-1.0pt}
\end{table}

\begin{figure}[h!]
\begin{center}
\includegraphics[totalheight=8cm,width=8cm]{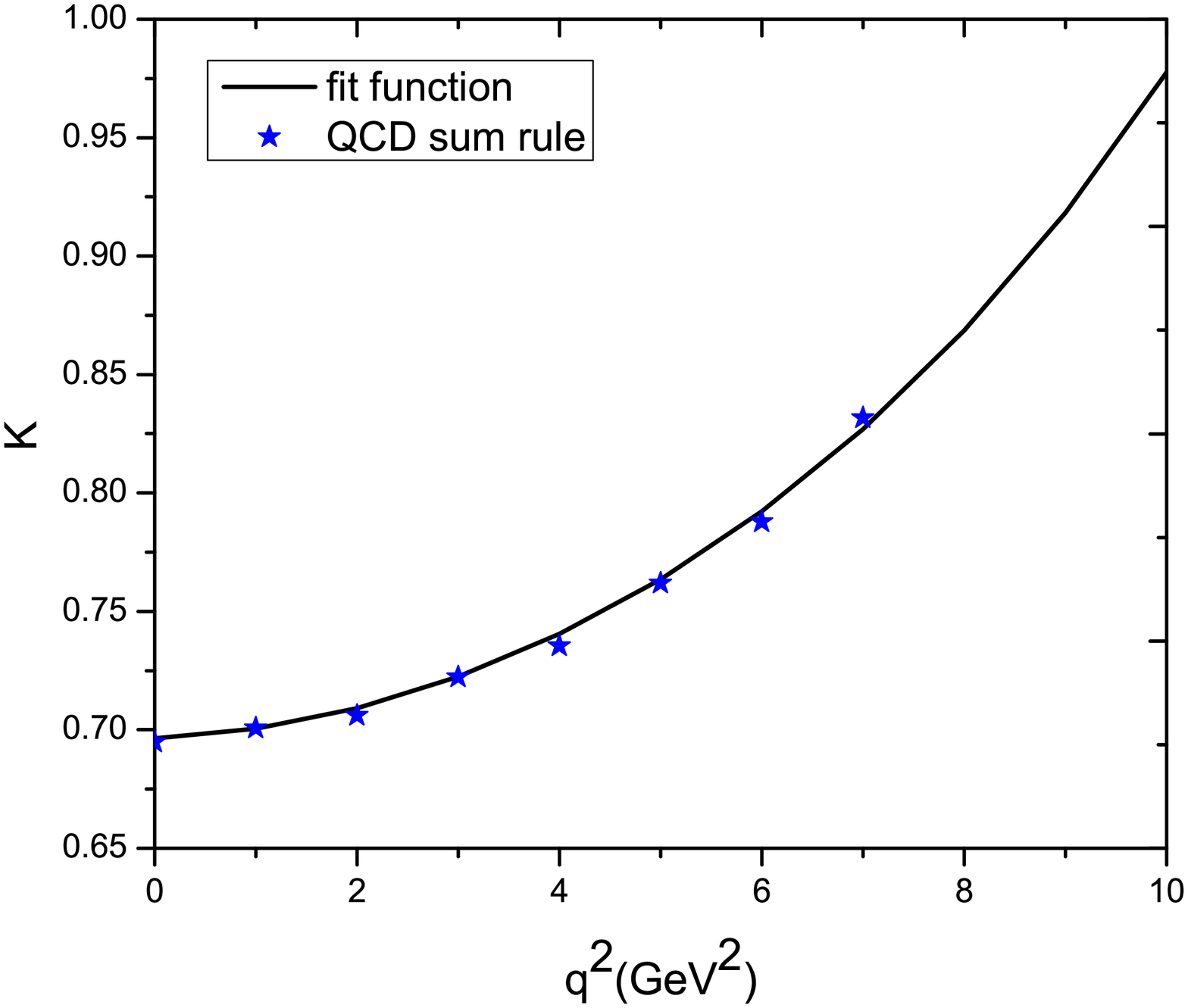}
\end{center}
\caption{\textbf K($q^2$) as a function of $q^2$ at
$M^2=15GeV^2$, $M^{\prime^2}=7.5GeV^2$, $s_0=35GeV^2$ and
$s_{0}^{\prime}=9GeV^2$.} \label{Kandhqsq}
\end{figure}

At this stage, we would like to find the behaviors of the
considered form factors in terms of $q^2$ using the  working
regions for the continuum thresholds and Borel mass parameters.
Our calculations depict that the form factors are truncated at
$q^2\simeq7~GeV^2$. To extend the results to the whole physical
region, we have to find a fit function such that
 it coincide with the QCD sum rules results at $q^2=(0-7)~GeV^2$ region.
Here, we should also stress that at the time-like momentum transfers   the spectral representations mainly develop  anomalous contributions, i.e.,
the double spectral densities receive contributions beyond those due to Landau-type singularities and deviate from
the corresponding Feynman amplitudes. This problem is discussed in details   in \cite{ball}. Although these contributions do not affect the values of the form factors at $q^2=0$ and
turn  out to be small at higher values of $q^2$ by the above-mentioned ranges of the auxiliary parameters in the decay channel under consideration, we take also into account these small contributions
in our numerical calculations. 
We find that  the form factors are well fitted to the following
function (see figure 3)\cite{Melikhov}:
\begin{eqnarray}\label{fitfunc}
f(q^2)=\frac{f_0}{\Big(1-\frac{q^2}{m_{B_s}^2}\Big)\Big[1-\sigma_1\Big(\frac{q^2}{m_{B_s}^2}\Big)
+\sigma_2\Big(\frac{q^2}{m_{B_s}^2}\Big)^2\Big]},
\end{eqnarray}
where the values of the parameters $f_0$, $\sigma_1$ and
$\sigma_2$, as an example  at $\mu=2~GeV$, are presented in table 2. The quoted errors in the
results are due to the errors in determinations of the working
regions of the continuum thresholds, Borel mass parameters as well
as uncertainties coming from other input
 parameters. Our numerical analysis show that setting $\mu$ from $2~GeV$ to $3~GeV$ increases the values of the form factors roughly with amount of 35\% at a fixed value of $q^2$. This rate of increase in the values of
the form factors are roughly  25\% when going from $\mu=3~GeV$ to $\mu=4~GeV$. These rates of changes reveal that the form factors depend on the scale parameter $\mu$, considerably.

In this part we would like to discuss the constraints that the HQET limit provides  on the form factors under discussion as the considered decay channel is based on the   heavy-to-heavy $b\to c$ transition at quark
level. Taking into account all the definitions and the values of the related parameters discussed in \cite{aka} (and references therein) for a similar channel, namely  $B_s\rightarrow D_{sJ}(2460)l \nu$,
we find that the HQET limit affects the
form factors $h(0)$ and  $K(0)$ more than the form factors $b_+(0)$ and $b_-(0)$ such that the values of the  form factors  $h(0)$ and  $K(0)$ decrease by 35\% and 42\%, respectively.
In contrast, the form factors $b_+(0)$ and $b_-(0)$  increase by 16\% and 5\%, respectively.

Having found the fit function of the  form factors in terms of $q^2$
at full physical region, now we  calculate the decay width of the
process under consideration. The differential decay width for
$B_{s} \rightarrow D_{s2}^{*}(2573)\ell\bar{\nu }_{\ell}$  transition is
obtained as  (see also \cite{numres})
\begin{eqnarray}\label{decaywidth}
\frac{d\Gamma}{dq^2}&=&\frac{\lambda(m_{B_{s}}^2,m_{D_{s2}^{*}}^2,q^2)}{4m_{D_{s2}^{*}}^2}
\Big(\frac{q^2-m_{\ell}^2}{q^2}\Big)^2\frac{\sqrt{\lambda(m_{B_{s}}^2,m_{D_{s2}^{*}}^2,q^2)}G_F^2
V_{cb}^2}{384m_{B_{s}}^3\pi^3}\Bigg\{\frac{1}{2q^2}\Bigg[3m_{\ell}^2\lambda(m_{B_{s}}^2,m_{D_{s2}^{*}}^2,q^2)
[V_0(q^2)]^2
\nonumber \\
&+&(m_{\ell}^2+2q^2)\Big|\frac{1}{2m_{D_{s2}^{*}}}\Big[(m_{B_{s}}^2-m_{D_{s2}^{*}}^2-q^2)
(m_{B_{s}}-m_{D_{s2}^{*}})V_1(q^2)
-\frac{\lambda(m_{B_{s}}^2,m_{D_{s2}^{*}}^2,q^2)}{m_{B_{s}}-m_{D_{s2}^{*}}}V_2(q^2)\Big]\Big|^2\Bigg]
\nonumber \\
&+&\frac{2}{3}(m_{\ell}^2+2q^2)\lambda(m_{B_{s}}^2,m_{D_{s2}^*}^2,q^2)\Bigg[\Big|
\frac{A(q^2)}{m_{B_{s}}-m_{D_{s2}^{*}}}-\frac{(m_{B_{s}}-m_{D_{s2}^{*}})V_1(q^2)}{\sqrt{\lambda(m_{B_{s}}^2,m_{D_{s2}^{*}}^2,q^2)}}
\Big|^2
\nonumber \\
&+&\Big|
\frac{A(q^2)}{m_{B_{s}}-m_{D_{s2}^{*}}}+\frac{(m_{B_{s}}-m_{D_{s2}^{*}})V_1(q^2)}{\sqrt{\lambda(m_{B_{s}}^2,m_{D_{s2}^{*}}^2,q^2)}}
\Big|^2\Bigg]\Bigg\},
\end{eqnarray}
where
\begin{eqnarray}\label{decaywidth1}
A(q^2)&=&-(m_{B_{s}}-m_{D_{s2}^{*}})h(q^2),
\nonumber \\
V_1(q^2)&=&-\frac{K(q^2)}{m_{B_{s}}-m_{D_{s2}^{*}}},
\nonumber \\
V_2(q^2)&=&(m_{B_{s}}-m_{D_{s2}^{*}})b_{+}(q^2),
\nonumber \\
V_0(q^2)&=&\frac{m_{B_{s}}-m_{D_{s2}^{*}}}{2m_{D_{s2}^{*}}}V_1(q^2)-\frac{m_{B_{s}}+m_{D_{s2}^{*}}}{2m_{D_{s2}^{*}}}V_2(q^2)
-\frac{q^2}{2m_{D_{s2}^{*}}}b_{-}(q^2), \nonumber\\
\lambda(a,b,c)&=&a^2+b^2+c^2-2ab-2ac-2bc.
\end{eqnarray}
Performing the integral over $q^2$ in the above equation at whole physical region, finally,  we obtain the values of the total decay widths and branching ratios for all lepton channels as 
 presented in tables 3, 4 and 5 for $\mu=2~GeV$, $\mu=3~GeV$ and $\mu=4~GeV$, respectively.
From these tables we see that when setting $\mu$ from $2~GeV$ to $3~GeV$, the decay rate and branching ratio increase by roughly 82\%, but when going from $\mu=3~GeV$ to $\mu=4~GeV$ the rate of increase
in these quantities is roughly 40\% for all lepton channels. From these changes,  we conclude that the results  of these quantities also depend considerably on the scale parameter $\mu$. 
The orders of branching fractions show that the semileptonic $ B_{s} \rightarrow D_{s2}^{*}(2573)\ell\overline{\nu }_{\ell}$ is accessible,  experimentally at all lepton channels
in near future.
\begin{table}[h]
\renewcommand{\arraystretch}{1.5}
\addtolength{\arraycolsep}{3pt}
$$
\begin{array}{|c|c|c|}
\hline \hline
     \mbox{  } &\Gamma(GeV) &   Br   \\
\hline
  \mbox{$B _s\rightarrow D_{s2}^{*}(2573)\tau\overline{\nu}_{\tau}$} &(2.82\pm1.32)\times 10^{-16}&
  (5.08\pm2.38)\times 10^{-4} \\
  \hline
  \mbox{$B _s \rightarrow D_{s2}^{*}(2573)\mu\overline{\nu}_{\mu}$} & (5.37\pm2.44)\times 10^{-16}&
   (1.19\pm0.54)\times10^{-3} \\
  \hline
  \mbox{$B _s \rightarrow D_{s2}^{*}(2573)e\overline{\nu}_{e}$} & (5.41\pm2.48)\times 10^{-16}&
   (1.21\pm0.55)\times 10^{-3} \\
                      \hline \hline

\end{array}
$$
\caption{Numerical results for the decay widths and branching ratios at
different lepton channels for $\mu=2~GeV$.} \label{numresult}
\renewcommand{\arraystretch}{1}
\addtolength{\arraycolsep}{-1.0pt}
\end{table}
\begin{table}[h]
\renewcommand{\arraystretch}{1.5}
\addtolength{\arraycolsep}{3pt}
$$
\begin{array}{|c|c|c|}
\hline \hline
     \mbox{  } &\Gamma(GeV) &   Br   \\
\hline
  \mbox{$B _s\rightarrow D_{s2}^{*}(2573)\tau\overline{\nu}_{\tau}$} &(5.14\pm2.46)\times 10^{-16}&
  (9.26\pm4.33)\times 10^{-4} \\
  \hline
  \mbox{$B _s \rightarrow D_{s2}^{*}(2573)\mu\overline{\nu}_{\mu}$} & (9.79\pm4.45)\times 10^{-16}&
   (2.18\pm0.98)\times10^{-3} \\
  \hline
  \mbox{$B _s \rightarrow D_{s2}^{*}(2573)e\overline{\nu}_{e}$} & (9.86\pm4.52)\times 10^{-16}&
   (2.20\pm0.92)\times 10^{-3} \\
                      \hline \hline

\end{array}
$$
\caption{Numerical results for the decay widths and branching
ratios at different lepton channels for $\mu=3~GeV$.}
\label{numresult}
\renewcommand{\arraystretch}{1}
\addtolength{\arraycolsep}{-1.0pt}
\end{table}
\begin{table}[h]
\renewcommand{\arraystretch}{1.5}
\addtolength{\arraycolsep}{3pt}
$$
\begin{array}{|c|c|c|}
\hline \hline
     \mbox{  } &\Gamma(GeV) &   Br   \\
\hline
  \mbox{$B _s\rightarrow D_{s2}^{*}(2573)\tau\overline{\nu}_{\tau}$} &(7.20\pm3.38)\times 10^{-16}&
  (1.30\pm0.61)\times 10^{-3} \\
  \hline
  \mbox{$B _s \rightarrow D_{s2}^{*}(2573)\mu\overline{\nu}_{\mu}$} & (1.37\pm0.62)\times 10^{-15}&
   (3.06\pm1.39)\times10^{-3} \\
  \hline
  \mbox{$B _s \rightarrow D_{s2}^{*}(2573)e\overline{\nu}_{e}$} & (1.39\pm0.64)\times 10^{-15}&
   (3.08\pm1.40)\times 10^{-3} \\
                      \hline \hline

\end{array}
$$
\caption{Numerical results for the decay widths and branching
ratios at different lepton channels for $\mu=4~GeV$.}
\label{numresult}
\renewcommand{\arraystretch}{1}
\addtolength{\arraycolsep}{-1.0pt}
\end{table}

In summary,  taking into account the perturbative ${\cal O}(\alpha_s)$  corrections we have calculated the transition form factors governing the semileptonic $ B_{s} \rightarrow
D_{s2}^{*}(2573)\ell\bar{\nu }_{\ell}$ transition at all lepton channels using an appreciate three-point correlation function. The fit functions of the form factors have  been used to estimate the corresponding
decay widths and branching ratios. The orders of branching ratios indicate that such channels contribute to the total width of the $B_s$ meson, considerably. We hope that it will
 be possible to study these channels at LHCb
in near future. Comparison of the future data with the theoretical results can help us in understanding the internal structure and nature of the
$D_{s2}^{*}(2573)$ charmed-strange tensor meson.

\subsection{ACKNOWLEDGEMENT}
This work has been supported by the Scientific and Technological Research Council of Turkey (TUBITAK) under
the research project 114F018.

\newpage

\section*{Appendix }

In this appendix, as an example, we briefly show how we calculate the perturbative
${\cal O}(\alpha_s)$ corrections for the structure $q_{\alpha}g_{\beta\mu}$, i.e.,  $\rho_{\alpha_{s_1}}(s,s',q^2)$. After performing the trace of
Eq. (\ref{rhoalfs}) and taking into account also the contributions of all diagrams in figure 1, we use the Feynman parametrization to perform 
the  four-$k$ and four-$k^{\prime}$ integrals. First we perform the four integral over $k$. Using the Feynman parametrization, as an example for diagram (a) in figure 1,  one can write 
\begin{eqnarray}\label{FeynmanPar}
\frac{1}{A_{1}^{a}A_{2}^{b}A_{3}^{c}A_{4}^{d}A_{5}^{e}}&=&\frac{\Gamma[a+b+c+d+e]}
{\Gamma[a]\Gamma[b]\Gamma[c]\Gamma[d]\Gamma[e]}\int_{0}^{1}dx\int_{0}^{1}dy\int_{0}^{1}dz
\int_{0}^{1}dt\int_{0}^{1}dt'
\nonumber\\
&\times& \frac{x^{a-1}y^{b-1}z^{c-1}t^{d-1}t'^{e-1}}
{\{xA_1+yA_2+zA_3+tA_4+t'A_5\}^{a+b+c+d+e}}\delta(x+y+z+t+t'-1), 
\nonumber \\
\end{eqnarray}
where $A_1=[(p'+k)^2-m_c^2]$, $A_2=[(p+k)^2-m_b^2]$, $A_3=[(k-k')^2-m_s^2]$, $A_4=[k^2-m_s^2]$, $A_5=k^{'2}$ and $a=b=c=d=e=1$ for diagram (a). The next step is to  
 perform the integral over $t'$ using the DiracDelta in Eq. (\ref{FeynmanPar}), rearrange the denaminator of the integrand on
the right-hand side of this equation and use the 
shift
\begin{eqnarray}
 k\rightarrow k-\frac{p y+p' x-k' z}{x+y+z+t},
\end{eqnarray}
to make the denaminator  full-squared in terms of $k$, i.e., in the form of  $k^2-\Delta$, where $\Delta$ is a function of $k'$, $p$, $p'$, $x$, $y$, $z$, $t$ and quark masses.

The integral over $k$ is performed via the following table of
$D-$dimensional integrals:
\begin{eqnarray}\label{Integrals}
\int\frac{d^Dk}{(2\pi)^D}\frac{1}{(k^2-\Delta)^n}&=&\frac{i(-1)^n}{(4\pi)^{D/2}}\frac{\Gamma[n-D/2]}
{\Gamma[n]}\Big(\frac{1}{\Delta}\Big)^{n-D/2},
\nonumber \\
\int\frac{d^Dk}{(2\pi)^D}\frac{k^2}{(k^2-\Delta)^n}&=&\frac{i(-1)^{n-1}}{(4\pi)^{D/2}}\frac{D}{2}
\frac{\Gamma[n-D/2-1]}
{\Gamma[n]}\Big(\frac{1}{\Delta}\Big)^{n-D/2-1},
\nonumber \\
\int\frac{d^Dk}{(2\pi)^D}\frac{k^{\mu}k^{\nu}}{(k^2-\Delta)^n}&=&\frac{i(-1)^{n-1}}{(4\pi)^{D/2}}
\frac{g^{\mu\nu}}{2} \frac{\Gamma[n-D/2-1]}
{\Gamma[n]}\Big(\frac{1}{\Delta}\Big)^{n-D/2-1},
\nonumber \\
\int\frac{d^Dk}{(2\pi)^D}\frac{(k^2)^2}{(k^2-\Delta)^n}&=&\frac{i(-1)^{n}}{(4\pi)^{D/2}}\frac{D(D+2)}{4}
\frac{\Gamma[n-D/2-2]}
{\Gamma[n]}\Big(\frac{1}{\Delta}\Big)^{n-D/2-2},
\nonumber \\
\int\frac{d^Dk}{(2\pi)^D}\frac{k^{\mu}k^{\nu}k^{\rho}k^{\sigma}}{(k^2-\Delta)^n}&=&\frac{i(-1)^{n}}
{(4\pi)^{D/2}} \frac{\Gamma[n-D/2-2]}
{\Gamma[n]}\Big(\frac{1}{\Delta}\Big)^{n-D/2-2} \nonumber \\
&\times&\frac{1}{4}(g^{\mu\nu}g^{\rho\sigma}+g^{\mu\rho}g^{\nu\sigma}+g^{\mu\sigma}g^{\nu\rho}).
\end{eqnarray}
Now, we proceed to perform the four-integral over $k^{\prime}$. Again we try to make the denaminator of the integrand of the integration over $k'$  full-squared in terms of $k'$ viz. $k^{'2}-\Delta'$,
with $\Delta'$ being a function of $p$, $p'$, $x$, $y$, $z$, $t$ and quark masses, by using the shift: 
\begin{eqnarray}
k'\rightarrow k'-\frac{(py+p'x)z}{x+y+z-t^2-z^2-(x+y)(x+y+z)-t(2x+2y+z-1)}.
\end{eqnarray}
In this step, the integral over $k'$ is performed again using
the above table of $D-$dimensional integrals. Now, we come back to the four dimensions. For the terms which converge
we  directly set $D=4$, but for those that  diverge by setting $D=4$, the following relation is used:
\begin{eqnarray}\label{Replace}
\frac{\Gamma[2-D/2]}{(4\pi)^{D/2}}\Big(\frac{1}{\Delta'}\Big)^{2-D/2}=\frac{1}{(4\pi)^2}
\Big(\frac{2}{\epsilon}-\log\Delta'-\gamma+\log(4\pi)+{\cal
O}(\epsilon)\Big),
\end{eqnarray}
with $\epsilon=4-D$ and $\Delta'$  is  negative. To obtain the imaginary part,  we use the following relation:
\begin{eqnarray}\label{Replace}
\log\Big[-|\Delta'|\Big]=\log\Big[e^{i\pi}|\Delta'|\Big]=i\pi+\log\Big[|\Delta'|\Big].
\end{eqnarray}
We do the similar calculations for all diagrams in figure 1.
As a result, we obtain
\begin{eqnarray}\label{alfscorrection}
\rho_{\alpha_{s_1}}&=&\alpha_{s}\int_{0}^{1}dx\int_{0}^{1-x}dy\int_{0}^{1-x-y}dz\int_{0}^{1-x-y-z}dt
\Bigg\{\frac{1}{4\pi^3\Lambda^5}\Bigg[-t(x+y+t-1)\Big((x+y+t-1)^2
\nonumber\\
&\times& (6m_c(t-3x+5y)+m_b(3t-5x+11y))+z(x+y+t-1)
\Big(m_b(6t-2x+14y-3)
\nonumber\\
&+&2m_s(5x-11y-3t)-6m_c(2x-6y-2t+1)\Big)+6z^2(x+y+t-1)(m_b-m_s+2m_c)
\nonumber\\
&+&3z^3(m_b+2m_c-2m_s)
\Big)\Theta\Big[L_{1}[s,s^{\prime},q^2,x,y,z,t]\Big]
\nonumber\\
&+&y(x+y+t-1)\Big((x+y+t-1)^2(24m_c(x-y)+m_s(t-7x+9y)) +z(x+y+t-1)
\nonumber\\
&\times&(24m_c(1-t-2y)-2m_b(t-7x+9y)+m_s(2x+18y+10t-2))-6z^2(x+y+t-1)
\nonumber\\
&\times& (3m_b+8m_c-3m_s)
-3z^3(6m_b+8m_c-3m_s)\Big)\Theta\Big[L_{2}[s,s^{\prime},q^2,x,y,z,t]\Big]
\nonumber\\
&+&x(x+y+t-1)\Big((x+y+t-1)^2(m_s-m_b)(5t-3x+13y)-z(x+y+t-1)
\nonumber\\
&\times&\Big(24m_c(t-x+3y)+(3+2t-6x+10y)(m_b-m_s)\Big)
+6z^2(x+y+t-1)
\nonumber\\
&\times&
(m_b+4m_c-m_s)+3z^3(m_b+8m_c-m_s)\Big)\Theta\Big[L_{3}[s,s^{\prime},q^2,x,y,z,t]\Big]
\Bigg]
\nonumber\\
&+&\int_{0}^{1-x-y-z-t}dw\Bigg[\frac{1}{4\pi^3\Lambda^{\prime^5}}\Bigg(m_b(t^2+w^^2-3x+3x^2+w(3x+y-1)+y+2xy-y^2
+z
\nonumber\\
&+& 3xz-yz-z^2+t(4x+z+w-1))(12+11t^2+11w^2-23x-23y+23w(x+y-1)
\nonumber\\
&+& 11(x+y)^2-23z+22wz+23z(x+y)+11z^2+t(22x+22y+23z+23w-23)) +m_c
\nonumber\\
&\times&
(4t^4+w^4+2(x+y-1)^2(11x^2+y(9-7y)+x(4y-15))+z(x+y-1)(51x^2+y
\nonumber\\
&\times& (57-37y)+x(14y-43)-18)+z^2(43+12x-44y)
(x+y-1)-2z^3(13x+15y-16)
\nonumber\\
&-& 7z^4+t^3(11w+34x-2y+7z-14)+w^2(13+40x^2-16y^2+3y-28yz+x(24y-6z
\nonumber\\
&-& 53)+2z(8-9z))+w(73x
-122x^2+55x^3-27y-56xy+77x^2y+66y^2-11xy^2-33y^3
\nonumber\\
&+& 2z(x+y-1)(15+26x-30y)-2z^2(21x+27y-28)-20z^3-6)+2w^3 (5x+3y+2z
\nonumber\\
&-& 4)+t^2(16+12w^2+78x^2+6y(3-5y)-23yz-16z^2+x(48y+65z-102)+w(77x
\nonumber\\
&-& 11y-4(7+z)))+t(8w^3+2(x+y-1) (3+35x^2+y(20-19y)+2x(8y-23))+w^2
\nonumber\\
&\times&
(52x-4y-12z-25)+z(109x^2+y(97-67y)+x(42y-94))+z^2(59-4x-60y)
\nonumber\\
&-& 28z^3+w(23+121x^2
+38y-55y^2+34z-64yz-48z^2+6x(11y+8z-25))))-m_s
\nonumber\\
&\times&
(2t^4+11w^4+(x+y-1)^2(8x^2+y(5-4y)+x(4y-11))+z(x+y-1)(56x^2-55x
\nonumber\\
&+& 21y +40xy-16y^2-5)+z^2(x+y-1)(17+81x-23y)+z^3(19+25x-19y)-7z^4
\nonumber\\
&+& t^3(14x+2y+20z+16w-7)+w^3(47x+3y+26z-25)+w(60x-3-109x^2 +52x^3
\nonumber\\
&-&
16y-70xy+84x^2y+39y^2+12xy^2-20y^3+16z(x+y-1)(10x-3y)+z^2(13+97x
\nonumber\\
&-& 35y)-10z^3)+t^2(8+27w^2+30x^^2-y(1+6y)-37z
+24yz+29z^2+w(84x+12y
\nonumber\\
&+& 56z-35)+x(24y+96z-41))+w^2(17+79x^2+8y-31z-(y+z)(25y-12z)+x
\nonumber\\
&\times& (54y+119z-96))+t(25w^3+26x^3-6y
+y^2(19-10y)-3)+12z-12y^2z+6z^2
\nonumber\\
&\times&
(y-2)+3z^3+w^2(106x+2y+53z-44)+x^2(42y+132z-61)+2x(19+3y(y-7)
\nonumber\\
&-& 7z+60yz+55z^2)+w(22 +120x^2-24y^2+y(4+8z)+24x(4y+9z-6)+31z^2
\nonumber\\
&-&
56z))\Bigg)\Theta\Big[L_{4}[s,s^{\prime},q^2,x,y,z,t,w]\Big]+\frac{1}{4\pi^3\Lambda^{\prime\prime^5}}\Bigg(m_b
(-6t^4+(x+y-1)^2(33x^2+y(7-11y)
\nonumber\\
&+&
x(22y-9))+z(x+y-1)(9+75x^2-y(13+29y)+x(46y-57))+z^2(24+81x^2+15x
\nonumber\\
&\times& (6y-7)+y(9y-65))+
3z^3(16x+12y-7)+6z^4+w^3(18x+6y+6z-1)-t^3(12x
\nonumber\\
&+& 24y+12z+18w-19)+t^2(37x^2-18w^2+y(55-35y-12z)+17z-w(6x+42y+18z
\nonumber\\
&-& 37)+x
(15+2y+24z)-20)+w^2(2+59x^2-13y^2+z(18z-23)+y(48z-5)+x(46y
\nonumber\\
&+& 84z-45))+t(7-6w^3+75x^3+w^2(17+24x-12y)+y(60y-29y^2-38))
+4z-2yz
\nonumber\\
&\times&
(5+13y)+z^2(48y-23)+12z^3+x^2(121y+118z-88)+x(6-28y+17y^2-90z
\nonumber\\
&+& 92yz+84z^2)+2w(48x^2+25y-24y^2-3z+18yz+9z^2
+3x(8y+18z-5))+w(75x^3
\nonumber\\
&-& 29y^3+y^2(38-4z)+2y(z-1)(4+39z)+x^2(121y+140z-110)+(z-1)(1-25z
\nonumber\\
&+& 18z^2)+x(17y^2+8y(17z-9)+6
(z-1)(19z-6))))+m_s(17t^4-5w^4-(x+y-1)^2
\nonumber\\
&\times&
(27x^2+10x(y-3)+18y-17y^2)+w^3(16-38x+6y-42z)-2z(x+y-1)(15+42x^2
\nonumber\\
&+& 13y(1-2y) +x(16y-57))-z^2(x+y-1)(111x-25y-87)-4z^3(21x+10y-21)
\nonumber\\
&-& 27z^4+t^3(8x+52y+24z+46w-52)+t^2(53+36w^3-67x^2-22w(4+x-5y)
+69y^2
\nonumber\\
&-& 122y+2x(7+y-34z)-20z+6wz+64yz-30z^2)+2w(3-49x+88x^2-42x^3+21y
\nonumber\\
&+& 38xy-58x^2y-50y^2+10xy^2+26y^3-4z(x+y-1)
(25x-9y-13)+y(30+28z)+2x
\nonumber\\
&\times&
(21y+80z-53))+2t(w^3-42x^3+(y-1)^2(26y-9)+x^2(77-58y-89z)+z(y-1)
\nonumber\\
&\times& (17+47y)+2z^2(29-7y)-32z^3
-2w^2(5+17x-16y+15z)+2x(8y+5y^2+53z
\nonumber\\
&-& 21yz-40z^2-13)+w(18-78x^2-76y+58y^2+48z+18yz-63z^2-2x(10y+57z
\nonumber\\
&-& 30))))+m_c\{10w^4
-8t^4+(x+y-1)^2(x^2-3y(y-4)-2x(y+18))+z(x+y-1)
\nonumber\\
&\times&
(36+38x^2+31y-18y^2-101x+20xy)+2z^2(50+33x^2-y(23+11y)+x(22y-83))
\nonumber\\
&+& z^3
(65x+29y-92)+28z^4+t^3(28-14w+13x-23y+4z)+w^3(39x+3y+58z-32)
\nonumber\\
&+& t^2(62x^2-26y^2+6w^2+58y-36z-17yz+60z^2+w(24+65x
-43y+66z)+x(36y
\nonumber\\
&+& 91z-62)-32)+w^2(34+64x^2+6y-24y^2-156z+35yz+114z^2+x(40y+143z
\nonumber\\
&-& 114))+w(38x^3-18y^3+y^2(51-46z)
+y(z-1)(21+61z)+x^2(58y+130z-137)
\nonumber\\
&+&
x(111+2y(y-43)-280z+84yz+169z^2)+2(z-1)(6-61z+47z^2))+t[12+22w^3
\nonumber\\
&+& 38x^3-47y+y^2
(53-18y)+68z+12yz(1-4y)+z^2(35y-156)+76z^3+w^2(91x-17y
\nonumber\\
&+& 120z-36)+x^2(58y+128z-135)+x(85+2y(y-41)-228z+80yz+143z^2) +2w
\nonumber\\
&\times& (1+63x^2+32y-25y^2-96z+9yz+87z^2+x(38y+117z-88))]\} \Bigg)
\nonumber\\
&\times& \Theta\Big[L_{5}[s,s^{\prime},q^2,x,y,z,t,w]\Big]
\nonumber\\
&-&\frac{1}{4\pi^3\Lambda^{\prime\prime\prime^5}}\Bigg(m_b(-11t^4-3(x^2+x(y-1)+y(y-1))(3+x(16+11x)
-5y-16xy+2y^2)
\nonumber\\
&+&
z(15x^3+2x(y-1)(22+29y)+x^3(62y-92)+y^2(23-12y)-4y-7)+z^2(20+xz^2(61
\nonumber\\
&+& 53x-10y)-17y) +z^3(12y-26x-19)+6z^4+3t^3(w+y+z-22x-1)+w^3(1+8x
\nonumber\\
&-& 6y+6z)+t^2(3w^2-132x^2+w(21x-16y+28z-16)+21x(y+z-1) -(13+19y
\nonumber\\
&-&
25z)(y+z-1))+w^2(31x^2+y(23-18y)+x(58y-10z-7)+z(18z-17))+w(15x^3
\nonumber\\
&+& 2x(5+46y-22z)(y+z-1)+x^2 (40y+84z-70)-(y+z-1)(1+y(18y-25)+z(19
\nonumber\\
&-& 18z)))+t(14w^3-110x^3+33x^2(y+z-1)-2x(y+z-1)(26+5y-39z)
+(y+z-1)^2
\nonumber\\
&\times&
(1+48y-20z)+w^2(34x+76y+8z-27)+w(33x^2+2(y+z-1)(55y-13z-6)+2x
\nonumber\\
&\times& (12y+56z-43))))+m_s(5t^4+5w^4
+3(x^2+x(y-1)+y(y-1))(9x^2+(y-1)(9y-21x
\nonumber\\
&-& 1))+2w^3(7x+21y-z-8)-2z(x^2(18x-43)-62x-9-17y+xy
(31x+58)+2y^2(29
\nonumber\\
&+&
2x)-32y^3)+z^2(10y(2+3y)-53x^2-53+2x(56y-79))+4z^3(13+16x-6y)-17z^4
\nonumber\\
&+& t^3(42x-8w
-8(y+z-1))-t^2(3w^3-96x^2-(7+19y-25z-52x)(y+z-1)+52wx
\nonumber\\
&+&2w(14z-8y-5))+w^2(17-31x^2-116y+20z+12
(8y-3z)(y+z)+x(92z-8y-58))
\nonumber\\
&+&
2w(x^2(32-20y-42z)-18x^3-x(y+z-1)(37+29y-71z)+(y+z-1)(3+43y^2+z
\nonumber\\
&\times& (21-23z)
+y(20z-49)))+2t(43x^3-4w^3+x(y+z-1)(20+5y-39z-40x)-(29y
\nonumber\\
&-& 21z+3)(y+z-1)^2+w^2(5-17x-37y+13z)-
w(40x^2+2(y+z-1)(1+31y-19z)
\nonumber\\
&+& x(12y+56z-37)))
)+m_c\{t^4-10w^4+36x+17x^2-52x^3-x^4+36y-47xy-41x^2y
\nonumber\\
&+& 52x^3y-100y^2-14xy^2+24x^2y^2+92y^3+25xy^3-28y^4
+z(52x^3-x(y-1)(y-123)
\nonumber\\
&+&
x^2(52y-45)-4(y-1)(19y^2-20y-3))+z^2(32+28x^2+x(138-77y)+12y(3-5y))
\nonumber\\
&-& z^3(51x+4 (7+y))+8z^4+t^2(y+z+w-1)(6w+84x+4y+8z-3)-w^3(13x+58y
\nonumber\\
&+& 22z-32)+2t^3(8w+x+8(y+z-1))+w^2(26x^2+156y+36z-
6(y+z)(19y+z)-x
\nonumber\\
&\times&
(y+77z-62))+w(52x^3+x(y+z-1)(85+37y-115z)+x^2(50y+54z-43)-2(y+z
\nonumber\\
&-& 1)(6+47y^2+z(5-7z)
+y(40z-61)))+t[5w^3-2x^3+w^2(2+32x+53y-23z)+120x^2
\nonumber\\
&\times& (y+z-1)+(y+z-1)^2(12+43y-33z)+4x(y+z-1)(7y+9z-5)+
w(120x^2+(y
\nonumber\\
&+&
z-1)(19+91y-61z)+x(60y+68z-52))]\}\Bigg)\Theta\Big[L_{6}[s,s^{\prime},q^2,x,y,z,t,w]\Big]
\Bigg]
\Bigg\}, \nonumber\\
\end{eqnarray}
where
\begin{eqnarray}\label{Ls}
L_{1}[s,s^{\prime},q^2,x,y,z,t]&=&\frac{(x+y+z+t)}{\Lambda^2}\Big[(x+y+t-1)\Big(-s^{\prime}xt
-sty-q^2xy+m_b^2y(x+t)
\nonumber\\
&+&m_b^2y^2+m_c^2x(x+y+t)\Big)+z(x+y+t-1)(m_c^2x-s^{\prime}x+m_b^2y-sy)
\nonumber\\
&+&z^2(m_c^2x-s^{\prime}x+m_b^2y-sy) +m_s^2(t+z)\Lambda\Big],
\nonumber\\
L_{2}[s,s^{\prime},q^2,x,y,z,t]&=&\frac{(x+y+z+t)}{\Lambda^2}\Big[(x+y+t-1)\Big(-s^{\prime}xt
-sty-q^2xy+m_b^2y(x+t)
\nonumber\\
&+&m_b^2y^2+m_c^2x(x+y+t)\Big)+z(x+y+t-1)(m_c^2x+m_b^2(x+2y+t)
\nonumber\\
&-&st-q^2x)+z^2(m_c^2x-st-q^2x+m_b^2(x+2y+t-1))+m_b^2z^3
+m_s^2t\Lambda\Big],
\nonumber\\
L_{3}[s,s^{\prime},q^2,x,y,z,t]&=&\frac{(x+y+z+t)}{\Lambda^2}\Big[(x+y+t-1)\Big(-s^{\prime}xt
-sty-q^2xy+m_b^2(x+y+t)\Big)
\nonumber\\
&-&z(x+y+t-1)(s^{\prime}t-ym_b^2+yq^2)-z^2(s^{\prime}t-ym_b^2+yq^2)+m_s^2t\Lambda
\nonumber\\
& +&m_c^2(x+z)\Lambda \Big],
\nonumber\\
L_{4}[s,s^{\prime},q^2,x,y,z,t,w]&=&\frac{(x+y+z+t+w)}{\Lambda^{\prime^2}}\Big[stx-st^2x+swx-stwx-sw^2x-stx^2
-swx^2
\nonumber\\
&-&
m_c^2ty+s^{\prime}ty+m_c^2t^2y-s^{\prime}t^2y-m_c^2wy+s^{\prime}wy+twy(m_c^2-s^{\prime})+m_c^2w^2y
\nonumber\\
&-&
s^{\prime}w^2y-m_c^2xy+q^2xy+2m_c^2txy-q^2txy-stxy-s^{\prime}txy+m_c^2wxy
\nonumber\\
&-& swxy-s^{\prime}wxy+m_c^2x^2y-q^2x^2y
-m_c^2y^2+2m_c^2ty^2-s^{\prime}ty^2+m_c^2wy^2
\nonumber\\
&-& s^{\prime}wy^2+2m_c^2xy^2-q^2xy^2+m_c^2y^3
+z(-s^{\prime}(t^2+w(w+y-1)+t(x+y
\nonumber\\
&+& w-1))-x(sw+q^2(x+y+t+w-1))+m_c^2(t^2+w^2+(x+2y)
\nonumber\\
&\times& (x+y-1)+w(x+3y-1)+t(2x+3y+w-1))) +z^2(-s^{\prime}(t+w)
\nonumber\\
&-& q^2x+m_c^2(x+2y+2w+t-1))+m_c^2z^3+m_s^2(t+w)\Lambda^{\prime}
+m_b^2x\Lambda^{\prime}\Big],
\nonumber\\
L_{5}[s,s^{\prime},q^2,x,y,z,t,w]&=&\frac{(x+y+z+t+w)}{\Lambda^{\prime\prime^2}}\Big[s^{\prime}tw
(1-t-w) -m_b^2tx+q^2tx+m_b^2t^2x-q^2t^2x
\nonumber\\
&-& m_b^2wx+swx+2m_b^2twx-q^2twx-stwx+m_b^2w^2x-sw^2x-m_b^2x^2
\nonumber\\
&+& m_b^2tx^2
-q^2tx^^2+m_b^2wx^2-swx^2+m_b^2x^3+s^{\prime}wy-s^{\prime}twy-s^{\prime}w^2y
\nonumber\\
&-& m_b^2xy+q^2xy+m_b^2txy
-q^2txy+m_b^2wxy-swxy-s^{\prime}wxy+2m_b^2x^2y
\nonumber\\
&-& q^2x^2y-s^{\prime}wy^2+m_b^2xy^2-q^2xy^2+\Big(q^2
(y(1-y)-t(x+w+t-1)
\nonumber\\
&-& y(x+w+t))-w(s(x+w+t-1)+s^{\prime}(y+t))+m_b^2z(t^2+w^2+(x
\nonumber\\
&+& y-1)(2x+y)+w(3x+y-1)+ t(3x+y+2w-1))+z^2(m_b^2(2x
\nonumber\\
&+& y+2w+2t-1)-q^2(t+y)-sw)+m_b^2z^3+m_s^2w\Lambda^{\prime\prime}
+m_c^2(y+t)\Lambda^{\prime\prime}\Big)\Big],
\nonumber\\
L_{6}[s,s^{\prime},q^2,x,y,z,t,w]&=&\frac{(x+y+z+t+w)}{\Lambda^{\prime\prime\prime^2}}\Big[s(-t^2(x+y)
-t(x^2+x(w+y-1)+y(w+y
\nonumber\\
&-& 1))-w(x^2+x(w+y-1)+y(w+y-1)))+z(-x(sw+q^2(x+w
\nonumber\\
&+& t-1))-y(s(t+w)+q^^2(x+w-1))-q^2y^2 -s^{\prime}(t^2+w(w+y-1)
\nonumber\\
&+&t(x+ y+w-1))+m_c^2(t^2+w^2+x(x-1)+y(x-1)+y^2+t(2x
\nonumber\\
&+&y+w-1)+ w(x+2y-1)))+z^2 (m_c^2(x+2y+2w+t-1)
\nonumber\\
&-&q^2(x+y)-s^{\prime}(t+w))
m_c^2z^3+m_s^2(t+w)\Lambda^{\prime\prime\prime}+m_b^2(x+y)
\Lambda^{\prime\prime\prime}\Big],
\end{eqnarray}
and
\begin{eqnarray}\label{Delta}
\Lambda&=&t^2-x-y-z(1-z)+(x+y)(x+y+z)+t(2x+2y+z-1),
\nonumber\\
\Lambda^{\prime}&=&t^2+w^2+z^2+(x+y)(x+y+z)+t(2x+2y+z+w-1)+w(x+y+2z-1)
\nonumber\\
&-& x-y-z,
\nonumber\\
\Lambda^{\prime\prime}&=&t^2+w^2+z^2+(x+y)(x+y+z)+t(x+y+2z+2w-1)+w(x+y+2z-1)
\nonumber\\
&-& x-y-z,
\nonumber\\
\Lambda^{\prime\prime\prime}&=&t^2+w^2+z^2+x^2+y^2+xy+z(x+2y-1)+t(2x+y+z+w-1)
\nonumber\\
&+&w(x+2y+2z-1)- x-y.
\end{eqnarray}

\end{document}